\documentclass [reprint,twocolumn,amsmath,amssymb,amsfonts,superscriptaddress]{revtex4}
\usepackage{graphicx,subfigure,color}
\usepackage[dvips]{epsfig}

\usepackage{verbatim}

\begin{document}

\title{Current-driven domain wall motion with spin Hall effect: Reduction of threshold current density}
\author{Jisu Ryu}
\affiliation{PCTP and Department of
Physics, Pohang University of Science and Technology, Pohang,
Kyungbuk 790-784, Korea}
\author{Kyung-Jin Lee}
\affiliation{Department of Materials Science and Engineering, Korea University, Seoul 136-701, Korea}
\affiliation{KU-KIST Graduate School of Converging Science and Technology, Seoul 136-713, Korea}
\author{Hyun-Woo Lee}
\email{hwl@postech.ac.kr} \affiliation{PCTP and Department of
Physics, Pohang University of Science and Technology, Pohang,
Kyungbuk 790-784, Korea}

\begin{abstract}
We theoretically study the current-driven domain wall motion in the presence of both the spin Hall effect and an extrinsic pinning potential. The spin Hall effect mainly affects the damping ratio of the domain wall precession in the pinning potential. When the pinning potential is not too strong, this results in a significant reduction of a threshold current density for the depinning of a domain wall with certain polarity. We also propose one way to distinguish the spin Hall effect induced spin-transfer torque from the one induced by the Rashba spin-orbit coupling experimentally.
\end{abstract}

\maketitle

An electric control of magnetization by the spin-transfer torque (STT)~\cite{berger1996, slonczewski1996} is under intense investigation due to its device application potentials.~\cite{parkin2008, allwood2005} An essential prerequisite for the STT is to generate a spin flow of electrons. A simple intuitive method is to utilize spin-polarized conduction electrons in a ferromagnetic layer (FM).~\cite{parkin2008, allwood2005, yamanouchi2004, klaui2005, yamaguchi2004} Recently, a very different way to generate a spin flow was proposed,~\cite{liu2011,liu2012prl, liu2012science, haazen2012} which utilizes a pure spin current generated by the spin Hall effect (SHE).~\cite{hirsch1999, zhang2000} When a charge current flows in a nonmagnetic layer (NM) with strong atomic spin-orbit coupling, the SHE generates a pure spin current in the direction perpendicular to the charge current. Thus, when a thin FM is deposited on the NM, the pure transverse spin current is injected to the FM and generates the STT. Recent experiments~\cite{liu2012prl, liu2012science, haazen2012} demonstrated that the STT generated by the spin current can be strong enough to switch the magnetization direction of the FM. The magnitude of the spin current is often parameterized by the spin Hall angle, denoting the ratio between the source charge current density and the resulting spin current density. The spin Hall angle was estimated to be $0.06$ for Pt~\cite{liu2012prl} and $-0.15$ for $\beta$-Ta.~\cite{liu2012science} Here, the opposite signs of the Pt and $\beta$-Ta spin Hall angles imply opposite spin direction of the spin current in the two cases.

The SHE-induced STT (SHE-STT) can modify the current-driven domain wall (DW) motion. An in-plane electric current in a FM/NM system can generate the DW motion through two mechanisms. In one mechanism, the spin-polarized charge current flowing in the FM generates adiabatic and nonadiabatic STTs,~\cite{tatara2004, zhang2004, thiaville2005} of which effect on the DW motion has been examined extensively. The other mechanism is the pure spin current injected from the NM due to the SHE. A recent theoretical study~\cite{smseo2012} examined the effect of the SHE-STT on the current-driven DW motion in an ideal situation without any pinning centers suppressing the DW motion. Such an ideal condition is rarely achieved in practical situations, however, and it is well known~\cite{tatara2006, jryu2009} that pinning centers may affect the DW motion considerably. Also from an application point of view, pinning centers are needed to achieve reliable control of DW locations in DW-based devices. Notches~\cite{klaui2005} or local modulation of magnetic properties~\cite{yamanouchi2004} are commonly suggested forms of pinning centers for device applications. In this Letter, we study the SHE-STT effects on the current-driven DW motion in the presence of an extrinsic pinning potential. We show that the SHE can significantly reduce the threshold current density to depin a DW from the pinning potential, thereby lowering the energy cost for the operation of DW-based devices.

\begin{figure}
\centering
{\includegraphics [width=0.45\textwidth]{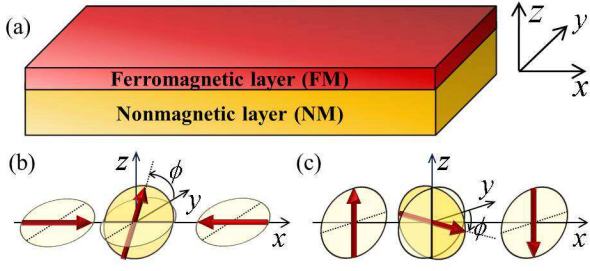}}
\caption{(color online) (a) Schematics of a FM/NM system. Magnetic configurations and definition of $\phi$ for a (b) N\'eel DW in IMA system (c) Bloch wall in PMA system. Red arrows represent magnetization direction.}\label{fig:wire}
\end{figure}

We consider a bi-layered system consisting of a thin FM deposited on a NM [Fig.~\ref{fig:wire}(a)]. The current-driven DW motion in this system can be described by generalizing the standard Landau-Lifshitz-Gilbert (LLG) equation~\cite{smseo2012, liu2011} to include the SHE-STT. In this study, we use the collective coordinate approach,~\cite{thiele1973, swjung2008} in which DW dynamics is described by two collective coordinates, DW position $q$ (along $x$-direction) and tilting angle $\phi$ [see Figs.~\ref{fig:wire}(b) and (c) for its definition in in-plane magnetic anisotropy (IMA) system and perpendicular magnetic anisotropy (PMA) system, respectively]. Although this approach has some limitations especially for vortex DWs, it was found to be a very useful tool~\cite{smseo2012} to analyze transverse DWs that we consider [Figs.~\ref{fig:wire}(b) and (c)]. Interestingly, for both IMA [Fig.~\ref{fig:wire}(b)] and PMA [Fig.~\ref{fig:wire}(c)] systems, this approach results in the identical equations of motion describing the coupled dynamics of $q$ and $\phi$,~\cite{swjung2008, smseo2012, tatara2006}
\begin{align}
\frac{\partial q}{\partial t}-\alpha\lambda\frac{\partial \phi}{\partial t}&=
\frac{\gamma \lambda H_d}{2}\sin2\phi-b_J,\label{eq:eomoriginal}\\
\alpha\frac{\partial q}{\partial t}+\lambda\frac{\partial \phi}{\partial t}&=
-b_J (\beta+B_{\rm SH}\lambda\sin\phi)-\frac{\gamma\lambda}{2M_S}\frac{\partial V_{\rm ext}}{\partial q},\nonumber
\end{align}
where $\alpha$ is the Gilbert damping constant, $\lambda$ is the DW width, $\gamma$ is the gyromagnetic ratio, $H_d$ is the hard-axis anisotropy field, $b_J= (\hbar\gamma P/2eM_S)J_F$ is the magnitude of the adiabatic STT in the velocity dimension, $\beta b_J$ is the magnitude of the nonadiabatic STT, $M_S$ is the saturation magnetization, $V_{\rm ext}$ is an extrinsic pinning potential, and $B_{\rm SH}=\pi\theta_{\rm SH}J_N/2t_FPJ_F$ represents the magnitude of the SHE-STT. Here, $P$ and $t_F$ are the thickness and spin polarization of the FM, $\theta_{\rm SH}$ is the spin Hall angle of the system, and $J_{N}(J_F)$ is the current density in the NM(FM). $V_{\rm ext}$ is assumed to be a finite-ranged harmonic potential placed at $q=0$: $V_{\rm ext}= (V_0/\xi^2) (q^2-\xi^2)\Theta(\xi-|q|)$ where $\Theta(x)$ is the unit step function centered at $x=0$. Here, $V_0$ and $\xi$ are the depth and range of the pinning potential. For analysis, it is convenient to rewrite Eq.~\eqref{eq:eomoriginal} in terms of dimensionless quantities as
\begin{align}
\dot X-\alpha\dot\phi=&\sin2\phi-J,\label{eq:eomdimless}\\
\alpha\dot X+\dot\phi=&-J (\beta+B_{\rm SH}\lambda\sin\phi)-VX\Theta(\frac{\xi}{\lambda}-|X|).\nonumber
\end{align}
Here, $X=q/\lambda$, $J=2b_J/\gamma H_d\lambda$, $V= (2\lambda/M_SH_d\xi^2)V_0$, and $\dot O= (2/\gamma H_d)\partial O/\partial t$.

\begin{figure}
\centering
{\includegraphics [width=0.45\textwidth]{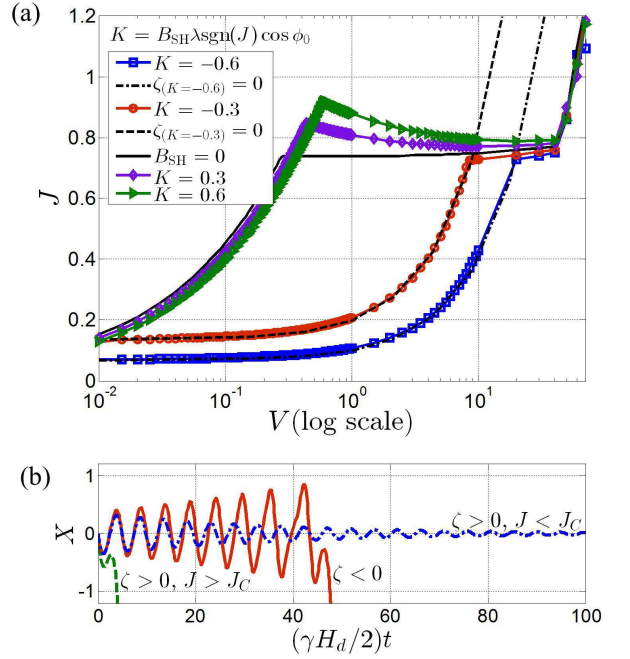}}
\caption{(color online) (a) $J_C$ as a function of $V$ for $B_{\rm SH}\lambda\cos\phi_0=-0.6$ (blue squares), $-0.3$ (red circles), $0$ (black solid line), $0.3$ (purple diamonds), and $0.6$ (green triangles). Solid lines connecting symbols are eye-guides. Black dashed and dashed-dotted lines represent $\zeta=0$ curves for $B_{\rm SH}\lambda\cos\phi_0=-0.3$ and $-0.6$, respectively. (b) $X$ after the current injection for $B_{\rm SH}=0$ and $J=0.5$ (blue dash-dotted line), $B_{\rm SH}=0$ and $J=0.8$ (green dashed line), and $B_{\rm SH}\lambda\cos\phi_0=-0.2$ and $J=0.5$ (red solid line). $\zeta=0.02$ for the first two lines and $\zeta=-0.013$ for the last line. }\label{fig:JcV}
\end{figure}

To investigate the SHE-STT effects in the presence of $V_{\rm ext}$, we first calculate the threshold current density $J_C$, above which a DW can get depinned from the pinning potential. In our simulation, a DW is placed at the center of the pinning potential ($X=0$) with the initial tilting angle (or polarity) $\phi_0=0$ or $\pi$. The current pulse is then turned on with zero rising time. Figure~\ref{fig:JcV}(a) shows the numerical simulation result obtained with $\alpha=0.02$, $\beta=0.01$ and $\xi/\lambda=1$. These parameter choices may be applicable to an IMA bi-layer system such as Pt/Py.~\cite{liu2011} In the absence of the SHE [black solid line in Fig.~\ref{fig:JcV}(a)], it is well known~\cite{tatara2006} that $J_C$ depends on $V$ in three distinct ways. In the so-called intermediate regime [$0.3<V<40$ in Fig.~\ref{fig:JcV}(a)], the adiabatic STT~\cite{tatara2006, swjung2008, koyama2011} is the main depinning mechanism and $J_C$ is almost independent of $V$, with $J_C$ value similar to the intrinsic threshold value.~\cite{tatara2004} For smaller (weak pinning regime, $V<0.3$) and larger (strong pinning regime, $V>40$) $V$, the nonadiabatic and adiabatic STTs are respectively the main depinning mechanisms overcoming the extrinsic pinning potential and $J_C$ increases as $V$ increases. A thick ($\sim 10$nm) Py layer~\cite{klaui2005, parkin2008, yamaguchi2004} can have small $V$ in the weak pinning regime. But when a Py layer is made thin ($\sim$ a few nm~\cite{liu2011}), which increases the SHE-STT since $B_{\rm SH}$ is inversely proportional to $t_F$, $V$ is expected to be larger due to the enhanced interface contribution to $V$. Thus a SHE bi-layer system is likely to be near the borderline between the weak and intermediate pinning regimes. Interestingly, it turns out that this is the range where the SHE-STT effects are most pronounced. Numerical simulation of Eq.~\eqref{eq:eomdimless} indicates a significant reduction of $J_C$ in this range if $B_{\rm SH}\lambda{\rm sgn}(J)\cos\phi_0<0$ [$-0.3$ and $-0.6$ for red circular and blue squared symbols in Fig.~\ref{fig:JcV}(a)]. On the other hand, $J_C$ changes only mildly in this range if $B_{\rm SH}\lambda{\rm sgn}(J)\cos\phi_0>0$ [$0.3$ and $0.6$ for purple diamond and green triangular symbols in Fig.~\ref{fig:JcV}(a)]. Note that for a given sign of $B_{\rm SH}$ (or $\theta_{\rm SH}$), the SHE reduces $J_C$ significantly only for a DW with a proper sign of $\cos\phi_0$ (or proper DW polarity). Hence, one can control the depinning efficiency by preparing the initial DW polarity to a proper value.

We remark that for metallic ferromagnets with IMA~\cite{liu2011}, our estimation of $|B_{\rm SH}\lambda|\sim 0.56$. Here, we assume $J_N=J_F$, $\theta_{\rm SH}=0.06$, $t_F=4{\rm nm}$, $P=0.7$, and $\lambda=20{\rm nm}$. 

\begin{figure}
\centering
{\includegraphics [width=0.45\textwidth]{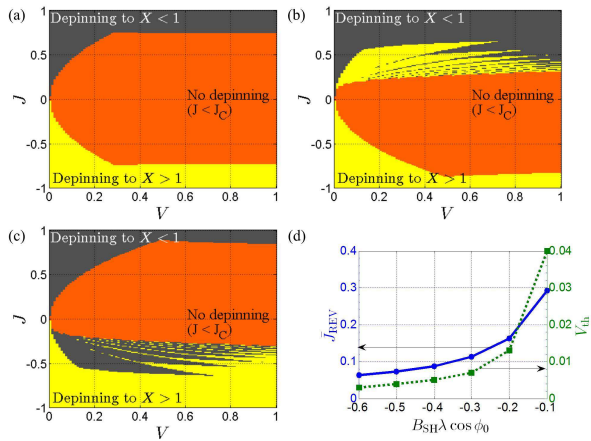}}
\caption{(color online) (a), (b), and (c): The DW depinning direction as a function of $J$ and $V$ for $B_{\rm SH}=0$, $B_{\rm SH}\lambda\cos\phi_0=-0.4$, and $B_{\rm SH}\lambda\cos\phi_0=0.4$, respectively. Orange, gray and yellow surfaces represent that the DW is not depinned, the DW depins to the left and right, respectively. For the reference, electrons flow to the left when $J>0$. (d) $\bar J_{\rm REV}$ (blue circles) and $V_{\rm th}$ (green squares) as a function of $B_{\rm SH}\lambda\cos\phi_0$. Solid and dotted lines connecting symbols are eye-guides. }\label{fig:depindirection}
\end{figure}

To understand the origin of the $J_C$ reduction, we further analyze $\phi$ dynamics. Within the pinning range [$\Theta(\xi/\lambda-|X|)=1$], two equations in Eq.~\eqref{eq:eomdimless} can be combined into the following $2$nd order partial differential equation for $\phi$ dynamics,
\begin{align}
 (1+\alpha^2)\ddot\phi+ (2\alpha\cos2\phi+B_{\rm SH}\lambda
J\cos\phi+\alpha V)\dot\phi&\label{eq:phi2ndPDE}\\
+ (\sin2\phi-J)V&=0.\nonumber
\end{align}
For small $J$, $\phi$ stays near its initial value $\phi_0$, which is either $0$ or $\pi$. Thus, to get an insight, we expand above equation near $\phi=\phi_0$ to obtain
\begin{align}
\ddot\phi+2\zeta\omega_0\dot\phi+\omega_0^2\phi=JV/ (1+\alpha^2),
\end{align}
where
\begin{align}
\centering
\omega_0^2=\frac{2 V}{1+\alpha^2}, ~~\zeta=\frac{\alpha (2+V)+B_{\rm SH}\lambda J\cos\phi_0}{\sqrt{8V (1+\alpha^2)}}.
\end{align}
Here, $\omega_0$ and $\zeta$ are an undamped angular frequency and the damping ratio, respectively. For $B_{\rm SH}=0$, $\zeta$ is always positive and $\phi$ approaches towards its stable steady value. The accompanied $X$ dynamics is shown in Fig.~\ref{fig:JcV}(b) as blue dash-dotted ($J<J_C$) and green dashed ($J>J_C$) lines. In the presence of the SHE ($B_{\rm SH}\neq 0$), however, $\zeta$ is not necessarily positive and may become {\it negative} if $B_{\rm SH}\lambda J\cos\phi_0$ is negative and sufficiently large in magnitude. When $\zeta<0$, $\phi$ dynamics does not have a stable steady value and results in the DW depinning driven by the amplification of $\phi$ [and also $X$ as shown in Fig.~\ref{fig:JcV}(b), red solid line]. Thus the condition $\zeta=0$ sets an upper bound on $|J_C|$, that is, $|J_C|<J_{C, {\rm up}}$ with
\begin{align}
\centering
J_{C, {\rm up}}\equiv\alpha(2+V)/|B_{\rm SH}\lambda|,\label{eq:Jcup}
\end{align}
which is shown as black dashed and dash-dotted lines in Fig.~\ref{fig:JcV}(a) for $B_{\rm SH}\lambda{\rm sgn}(J)\cos\phi_0=-0.3$ and $-0.6$, respectively. In the region where the SHE reduces $J_C$, note that $J_{C, {\rm up}}$ is in good agreement with numerically calculated $J_C$ values [red circles and blue squares in Fig.~\ref{fig:JcV}(a)]. This demonstrates that the reduction of $J_C$ by the SHE arises because the SHE can induce negative $\zeta$ and thus amplify the $\phi$ dynamics. For $B_{\rm SH}\lambda{\rm sgn}(J)\cos\phi_0>0$, on the other hand, the SHE increases $\zeta$ and thus $J_C$ as well [purple diamond and green triangular symbols in Fig.~\ref{fig:JcV}(a)]. A slight decrease of $J_C$ in the weak pinning regime is due to the shift of the stable DW position. Note that one sign of $B_{\rm SH}\lambda J\cos\phi_0$ reduces $\zeta$ while the other increases it. Thus, to reduce $J_C$, the DW polarity should be properly adjusted before the current injection, for example by applying weak in-plane magnetic field or injecting a properly designed train of small current pulses.~\cite{thomas2006}

Similar reduction of $J_C$ occurs in a PMA bi-layer as well, which is natural since IMA and PMA bi-layer systems share the same equations of motion for the DW depinning dynamics [Eq.~\eqref{eq:eomoriginal}]. In particular, the upper bound $J_{C, {\rm up}}$ in Eq.~\eqref{eq:Jcup} applies equally to the PMA bi-layer. Thus the difference between the PMA and IMA bi-layers may come only through different material parameters. For the PMA bi-layer system $\beta$-Ta/CoFeB with low Gilbert damping $\alpha=0.008$~\cite{liu2012science}, $J_{C, {\rm up}}$ is estimated to be similar to or even smaller than the corresponding value in Fig.~\ref{fig:JcV}(a) for the IMA bi-layer system. For the PMA bi-layer system Pt/CoFeB, the reported value of $\alpha$ ranges from $0.025$~\cite{liu2012science} to 0.5~\cite{miron2011}. Since $J_{C, {\rm up}}$ is proportional to $\alpha$, the reduction of $J_C$ due to the SHE is expected to be much less significant for the PMA system with large $\alpha$.

Next, we examine the depinning direction. In the absence of the SHE ($B_{\rm SH}=0$), the DW gets depinned to the right (left) of the pinning center when electrons flow to the right (left) and thus the depinning direction agrees with the electron flow direction [Fig.~\ref{fig:depindirection}(a)]. This direction remains the same even in the presence of the SHE if $B_{\rm SH}\lambda J\cos\phi_0>0$. On the other hand, when $B_{\rm SH}\lambda J\cos\phi_0<0$ and $\zeta<0$, interestingly, the depinning direction is {\it opposite} to the electron flow direction in a wide $J$ range near $J_C$ [see upper yellow region in Fig.~\ref{fig:depindirection}(b) and lower gray region in Fig.~\ref{fig:depindirection}(c)].

This situation resembles the situation for the DW motion direction in an ideal wire without pinning centers. In the absence of the SHE, the terminal velocity of the DW is always along the electron flow direction. When the SHE exists, however, it was recently demonstrated~\cite{smseo2012} that the terminal motion direction can be opposite to the electron flow direction if $B_{\rm SH}\lambda J\cos\phi<0$. Note that for both the depinning direction reversal and the terminal motion direction reversal, $B_{\rm SH}\lambda J\cos\phi$ should be negative. Thus when the initial DW polarity $\phi_0$ satisfies this condition for a given sign of $B_{\rm SH}\lambda J$, the DW may get depinned against the electron flow direction and maintain the same motion direction. This raises a possibility that a DW may propagate against the electron flow direction without getting pinned in a {\it nonideal} nanowire containing pinning centers. This may also provide an explanation for the reversed DW motion observed experimentally,~\cite{miron2011} without resorting to the Rashba spin-orbit coupling (RSOC) effect.~\cite{kwkim2012} For this possibility to work, it is important to avoid the DW polarity switching or the sign change of $\cos\phi$, which sets a constraint; $J$ should be smaller than the certain current density value ($\bar J_{\rm REV}$), above which the DW polarity switching occurs. Thus, $J$ should fall in the current density window, $J_C<J<\bar J_{\rm REV}$. For such window to exist, the inequality $J_C<\bar J_{\rm REV}$ should be satisfied. By numerical simulation, we calculate the threshold pinning potential depth $V_{\rm th}$, above which the inequality is violated. Firstly, $\bar J_{\rm REV}$ for $V=0$ [blue circular symbols in Fig.~\ref{fig:depindirection}(d)] is calculated. For $|B_{\rm SH}\lambda|=0.1\sim0.6$, $\bar J_{\rm REV}$ is of the order of $10^{-1}$ which amounts to the current density of the order of $10^{12}{\rm A/m}^2$ for $M_S=10^6{\rm A/m}$ and $H_d=1{\rm T}$. Then $\bar J_{\rm REV}$ is compared with $J_C$ to extract $V_{\rm th}$ [green squared symbols in Fig.~\ref{fig:depindirection}(d)]. For $|B_{\rm SH}\lambda|=0.1\sim0.6$, $V_{\rm th}$ is smaller than $0.04$ which amounts to $4\%$ of the hard-axis anisotropy energy of a DW.~\cite{swjung2008} Here, we used $V=(V_0\lambda/K_d\xi^2)$ and $\lambda\sim \xi$ where $K_d=M_S H_d/2$ is the hard-axis anisotropy energy. To observe the DW motion against the electron flow direction, $V<V_{\rm th}$ should be satisfied. Thus to explain the reversed DW motion in a recent experiment~\cite{miron2011} with the SHE-STT (but without the RSOC effect~\cite{kwkim2012}), pinning centers should be very weak.

Lastly, recent theory~\cite{kwkim2012} indicates that the RSOC-induced STT (RSOC-STT) may drive the DW against the electron flow direction in a FM sandwiched by two dissimilar layers such as metal and oxide. In this case, the so-calle field-like STT arising from the RSOC suppresses the DW polarity switching. Thus the DW motion against the electron flow direction can be maintained for relatively larger value of $V$ compared to the case with the SHE-only (without RSOC-STT) case. The strength of the extrinsic pinning potential can be controlled, for example by introducing the notch in the wire. Thus this feature may be used to distinguish the RSOC-STT and the SHE-STT in experiments.

In summary, we examine the SHE-STT effects on a DW motion in the presence of the extrinsic pinning potential. The presence of the SHE-STT mainly affects the damping ratio of a DW precession in the pinning potential and by inducing negative damping ratio, it can significantly reduce the threshold current density. We also examined the DW motion direction and found that the SHE-STT can induce a DW motion against the electron flow direction if the pinning potential is sufficiently weak. By using this dependence on the pinning potential strength, we suggest one way to distinguish the SHE-STT and RSOC-STT in the presence of the pinning potential.

\end{document}